\documentclass[12pt]{article}
\usepackage{amssymb}
\usepackage[dvips]{epsfig}

\newcommand{\be}{\begin{equation}}
\newcommand{\ee}{\end{equation}}
\def\bea{\begin{eqnarray}}
\def\eea{\end{eqnarray}}

\newcommand{\bn}{\begin{eqnarray}}
\newcommand{\en}{\end{eqnarray}}

\newcommand{\p}{\partial}

\newcommand{\nn}{\nonumber}

\newcommand{\no}{\noindent}

\newcommand{\s}{\,\,\,\,}

\def\bea{\begin{eqnarray}}
\def\eea{\end{eqnarray}}

\newcommand{\beq}{\begin{eqnarray}}
\newcommand{\eeq}{\end{eqnarray}}
\begin{document}

\title{\textbf{Dual descriptions of massive spin-3 particles in $D=2+1$ via Noether gauge embedment}}
\author{E.L. Mendon\c ca\footnote{eliasleite@feg.unesp.com}, D. Dalmazi\footnote{dalmazi@feg.unesp.br},
\\
\textit{{UNESP - Campus de Guaratinguet\'a - DFQ} }\\
\textit{{Avenida Dr. Ariberto Pereira da Cunha, 333} }\\
\textit{{CEP 12516-410 - Guaratinguet\'a - SP - Brazil.} }\\}
\date{\today}
\maketitle

\begin{abstract}

 We present here a relationship among massive self-dual models for spin-3 particles in $D=2+1$ via the Noether Gauge Embedment $(NGE)$ procedure. Starting with a first order model (in derivatives) $S_{SD(1)}$ we have obtained a sequence of four self-dual models $S_{SD(i)}$ where $i=1,2,3,4$. We demonstrate that the $NGE$ procedure generate the correct action for the auxiliary fields automatically.
 We obtain the whole action for the  $4th$ order self-dual model including all the needed auxiliary fields to get rid of the ghosts of the theory. 

\end{abstract}

\newpage

\section{ Introduction}

Massive gauge field theories in three spacetime dimensions has attracted 
much attention since long ago. A special feature of such theories
is that they can be massive without gauge symmetry breaking. This is possible
thanks to the addition of topological terms, as it was done in the case of 
topologically massive electrodynamics TME and topologically massive gravity 
TMG \cite{DJT}. The two models cited have in common the fact that they describe only singlets of 
massive spin-2 particles i.e. describe only one helicity
mode $+s$ or $-s$, where we have $s=1$ for TME and $s=2$ for TMG. In the spin-2 case
however through the Noether Gauge Embedment NGE procedure we have demonstrated \cite{ddelm1} that the TMG
actually can be related, at least at the level of equations of motion, to other 
three models forming a sequence of the so called self-dual 
models of first, second, third (TMG) and fourth 
order in derivatives. 

With these self-dual models we have demonstrated \cite{ddelm3,ddelm4} that at the linearized level the Fierz-Pauli action which describes a doublet of massive spin-2 particles can be obtained via a soldering procedure of two second order self-dual models of opposite helicities. Besides, one can recover the New Massive Gravity NMG \cite{Tow1} (also at the linearized level) by soldering two self-dual models of opposite helicities of either third or fourth order in derivatives. 

Although we do not have observed higher spins particles ($s\geqq 2$) in nature, the string theory predicts the existence of such particles, so it would be interesting to investigate if the same duality procedure can be generalized for such particles in the context of gauge theories .
A way of introducing this context is starting by the models of spin-3. That is only because 
this is the simplest higher spin theory we can deal with. Strictly speaking the spin-3 theories do not contain for example the double trace condition present in the context of higher spin theories in general ($s>3$). In the spin-2 case, as mentioned here one can observe that we have four self-dual models in $D=2+1$. In the spin-3 case there are apparently six self-dual models \cite{tow3}. There seems to be a rule of $2s$ for the number 
of models according to the spin-s we have.

The first two models which describe one massive spin-3 mode in $D=2+1$ were proposed by  Aragone and Khoudeir in \cite{AragoneS31} and \cite{AragoneS32}. In both models the authors make use of the vierbein formulation which was introduced by Vasiliev in \cite{Vasiliev}. A Chern-Simons like term is present in both the first and the second order self-dual model. In the second order self-dual model the kernel of the action is the usual massless spin-3 second order term. The third model that we have in sequence is named topological massive spin-3 theory in analogy with the spins one and two cases. The formulation of this model however is made in terms of totally symmetric fields which indicate that, in trying to connect those models via the NGE procedure one needs to deal with a change of variables which relates the non-symmetric formulation in terms of vierbeins and the totally symmetric one.

It is a feature of massive models in $D=2+1$ that in order to have only the healthy spin-3 mode one needs to add an auxiliary Lagrangian which contains ghost-killing fields. In the three models cited above besides the healthy mode we have ghosts of spin-1. That is why in those models we have always an auxiliary Lagrangian which has the role of eliminating the spurious degrees of freedom. In \cite{Tow1} the authors propose a fourth order equation of motion for a massive spin-3 particle, although they do not propose an action they suspect that it should contain auxiliary fields. Here we start with the first order self-dual spin-3 model of \cite{AragoneS31}, which we call $SD_1^{(3)}$ and obtain a sequence of higher order self-dual models: $SD_i^{(3)}\to SD_{i+1}^{(3)}$ with $i=1,2,3$ via NGE. At each stage a new local symmetry is added via embedding. In all cases the full action, including auxiliary fields, is presented. There is no need of fine tuning the ``ghost killing'' auxiliary action. The whole models $SD_i^{(3)}$, $i=2,3,4$ follow from the first order model. However we have not been able to obtain the fifth and the sixth order self-dual models of \cite{Tow1}. 

The paper is organized as follows, in the first section we present the first order self-dual model and the embedment of the first symmetry. In the second section we make the $NGE$ procedure to the second order self-dual model. In the third section we rewrite our third order self-dual model in  terms of totally symmetric fields and make the last immersion leading to a fourth-order self-dual model.

\section{NGE procedure for the spin-3 self-dual models}

\subsection{The starting point}
\subparagraph{}
 
We begin with the first order self-dual model for a massive spin-3 particle in $D=2+1$ dimensions given by the action below: 

\bea S_{SD(1)}[\omega, A]&=&\int d^3x\left\lbrack -\frac{m}{2}\xi_{\mu(\beta\gamma)}\omega^{\mu(\beta\gamma)}+\frac{m^2}{6}(\omega_{\mu}\omega^{\mu}-\omega_{\mu(\beta\gamma)}\omega^{\beta(\mu\gamma)})+m^2\omega_{\mu}A^{\mu}+j_{\mu(\beta\gamma)}\omega^{\mu(\beta\gamma)} \right\rbrack\nn\\
 &+&  S^{(1)}[A]. \label{sd1}\eea

\no For convenience it is useful to define the notation \be \xi_{\mu(\beta\gamma)}=E_{\mu}^{\s\lambda}\omega_{\lambda(\beta\gamma)}\ee

\no where $E_{\mu\nu}=\epsilon_{\mu\nu\gamma}\p^{\gamma}$. The action (\ref{sd1}) is proposed by \cite{AragoneS31} in the vierbein formulation. In the context of higher spin particles, this formulation was first introduced by Vasiliev in \cite{Vasiliev}. Here, we work in a flat spacetime with signature $(-,+,+)$ and the basic spin-3 field is $\omega_{\mu(\beta\gamma)}$, which is symmetric and traceless in its Lorentz-like indices i.e $\omega_{\mu(\beta\gamma)}=\omega_{\mu(\gamma\beta)}$ with $\eta^{\beta\gamma}\omega_{\mu(\beta\gamma)}=0$. We define $\omega_{\gamma}=\eta^{\mu\beta}\omega_{\mu(\beta\gamma)}$.  Then, one can identify the first term in (\ref{sd1}) as a first order Chern-Simons like term. The mass term is similar to the Fierz-Pauli mass term for spin-2 particles, except for the extra term $m^2\omega_{\mu}A^{\mu}$ which includes the auxiliary vector field $A_{\mu}$.

Massive spin-3 actions can not avoid the presence of auxiliary fields, see for example section-3 of \cite{AragoneDeserYang}. In (\ref{sd1}) besides the spin-3 propagation one also has a residual spin-1 mode, and that is why we have the action $S^{(1)}[A]$ which contains dynamic terms for the ghost-killing field of spin one, and it is given \cite{AragoneS31} by:
 
 \be S^{(1)}[A]=\int d^3x\,\,\left\lbrack-9m\,\,\epsilon^{\mu\nu\alpha}A_{\mu}\p_{\nu}A_{\alpha}-9m^2\,\,A_{\mu}A^{\mu}
 \,-12(\p_{\mu}A^{\mu})^2 \right\rbrack.\label{lagaux1}\ee  
 
\no As we write down the action (\ref{sd1}) with a spin-3 sector governed by $\omega_{\mu(\beta\gamma)}$ and a spin-1 sector governed by $A_{\mu}$ one might wonder what would be the simplest contact term between these two sectors, and this is of course given by the term $\omega_{\mu}A^{\mu}$. Finally in order to have dual maps between the self-dual models we have added a source term to the spin-3 sector given by $j_{\mu(\beta\gamma)}\omega^{\mu(\beta\gamma)}$, the source $j_{\mu(\beta\gamma)}$ shares the same indices symmetry of $\omega_{\mu(\beta\gamma)}$. 

Due the presence of the mass term in the spin-3 sector, the gauge symmetry of the Chern-Simons like term:

\be \delta_{\tilde{\Lambda}} \omega_{\mu(\beta\gamma)}= \p_{\mu}\tilde{\Lambda}_{(\beta\gamma)},\label{gauge1}\ee

\no is broken. Where we have a traceless parameter  $\eta^{\beta\gamma}\tilde{\Lambda}_{(\beta\gamma)}=0$ . As we have done for spin-2 particles in $D=2+1$ dimensions, we would like to systematically impose  this gauge symmetry by using the $NGE$ procedure in order to obtain an invariant model, with the same particle content of the first order self-dual model (\ref{sd1}).  Our previous experience with spin-2 self-dual models tells us that the price one pays in obtaining an invariant model is that this new model is of higher order in derivatives. 
 
The $NGE$ procedure consists of modifying the original action, which is non invariant under (\ref{gauge1}), adding a quadratic term in the Euler tensor. By this way we are automatically ensuring that the equations of motion of the original model (\ref{sd1}) are embedded in the new model. First let us write the Euler tensor which comes from the spin-3 sector, by taking the variation of the action with respect to $\omega_{\mu(\beta\gamma)}$:

\be \delta S_{SD(1)} = \int d^3x\s K^{\mu(\beta\gamma)}\,\,\delta\omega_{\mu(\beta\gamma)},\ee

\no The Euler tensor is given by:

\be K^{\mu(\beta\gamma)}=-m\xi^{\mu(\beta\gamma)}+\frac{m^2}{6}(\eta^{\mu\beta}\omega^{\gamma}+\eta^{\mu\gamma}
\omega^{\beta}-\omega^{\beta(\mu\gamma)}-\omega^{\gamma(\mu\beta)})
+\frac{m^2}{2}f^{\mu(\beta\gamma)}(A)+j^{\mu(\beta\gamma)}.\label{euler1}\ee

\no One can notice that $K^{\mu(\beta\gamma)}=K^{\mu(\gamma\beta)}$ and $\eta_{\beta\gamma}K^{\mu(\beta\gamma)}=0$. Besides,  we have defined:

\be f^{\mu(\beta\gamma)}(A)=\eta^{\beta\mu}A^{\gamma}+\eta^{\gamma\mu}A^{\beta}-\frac{2}{3}\eta^{\beta\gamma}A^{\mu}.\label{fa}\ee 

\no The next step in the $NGE$ procedure consists basically of performing two iterations, the first one can be done by coupling a new auxiliary tensor field $a_{\mu(\beta\gamma)}$ to the Euler tensor as follow: 

\be S_{1}= S_{SD(1)}-\int\,d^3x\,a_{\mu(\beta\gamma)}\,K^{\mu(\beta\gamma)}.\label{first}\ee
 
\no Now let us take the gauge variation of (\ref{first}) by choosing a proper gauge variation for the auxiliary field $a_{\mu(\beta\gamma)}$ such as $\delta_{\tilde{\Lambda}}\omega_{\mu(\beta\gamma)}=\delta_{\tilde{\Lambda}}a_{\mu(\beta\gamma)}=\p_{\mu}\tilde{\Lambda}_{(\beta\gamma)}$. Then we have:
  
\be  \delta_{\tilde{\Lambda}}S_1=-\int\,d^3x\,\delta_{\tilde{\Lambda}}K^{\mu(\beta\gamma)}a_{\mu(\beta\gamma)}.\label{var28}\ee

\no Taking the gauge variation of the Euler tensor (\ref{euler1}) with $\delta_{\tilde{\Lambda}}A^{\mu}=0$ and substituting back in (\ref{var28}), we end up with: 
\be S_2=S_1-\int\,d^3x\,\left\lbrack a_{\mu(\beta\gamma)}K^{\mu(\beta\gamma)}-\frac{m^2}{6}(a_{\mu}a^{\mu}-a_{\mu(\beta\gamma)}a^{\beta(\mu\gamma)})\right\rbrack,\ee

\no after eliminating the auxiliary field $a_{\mu(\beta\gamma)}$ through its algebraic equations of motion to the following second iterated action:

\be S_2=S_{SD(1)}-\frac{3}{2m^2}\int d^3x\,\, \left\lbrack 2K_{\mu(\beta\gamma)}K^{\beta(\mu\gamma)}-K_{\mu(\beta\gamma)}K^{\mu(\beta\gamma)}\right\rbrack.\label{final1}\ee

As it was expected the action (\ref{final1}) is quadratic on the Euler tensor which ensures that the equations of motion of (\ref{sd1}) $K^{\mu(\beta\gamma)}=0$ are embedded in the equations of motion of (\ref{final1}). By construction, $S_2$ is automatically gauge invariant under (\ref{gauge1}). After substituting back the Euler tensor (\ref{euler1}) in (\ref{final1}) we obtain the second order spin-3 self-dual model:

\be S_{SD(2)}=\int d^3x\,\,\left\lbrack \frac{1}{2}\xi_{\mu(\beta\gamma)}\Omega^{\mu(\beta\gamma)}(\xi)+\frac{m}{2}\xi_{\mu(\beta\gamma)}\omega^{\mu(\beta\gamma)}+2m\xi_{\mu}A^{\mu}-j_{\mu(\beta\gamma)}F^{\mu(\beta\gamma)}(\omega,A)\right\rbrack
 + S^{(2)}[A].\label{EHCS1}\ee

\no The first term in (\ref{EHCS1}) is the usual massless spin-3 second order action  in the vierbein-like formulation, where:

\be \Omega_{\mu(\beta\gamma)}(\xi)= 3(\xi_{\beta(\mu\gamma)}+\xi_{\gamma(\mu\beta)}-\xi_{\mu(\beta\gamma)})-2\eta_{\beta\gamma}\xi_{\mu},\ee

\no The symbol $\Omega$ is self-adjoint: \be \int d^3x\,\omega_{\mu(\beta\gamma)}\Omega^{\mu(\beta\gamma)}(\xi)=\int d^3x\,\xi_{\mu(\beta\gamma)}\Omega^{\mu(\beta\gamma)}(\omega).\ee
The second order term comes out as an analogue of the Einstein-Hilbert term for spin-2. Actually as we are going to verify ahead it is possible to rewrite it in terms of a totally symmetric field $\phi_{\mu\beta\gamma}$. The gauge invariant second order action (\ref{EHCS1}) has now a new auxilary action $S^{(2)}[A]$ given by:
 
 \be S^{(2)}[A]=\int d^3x\,\,\left\lbrack-9m\,\,\epsilon^{\mu\nu\alpha}A_{\mu}\p_{\nu}A_{\alpha}-\frac{32m^2}{3}\,\,A_{\mu}A^{\mu}
 \,-12(\p_{\mu}A^{\mu})^2 \right\rbrack,\label{lagaux2}\ee

\no which differs from (\ref{lagaux1}) just by a numerical factor on the Proca mass term. It is not difficult to show after a rescaling of $A_{\mu}$ that the action (\ref{EHCS1}) is in fact precisely the second order self dual model proposed by \cite{AragoneS32}.  Another difference has automatically appeared, as one can see the linking term of the spin-3 field with the spin-1 auxiliary field has changed from  $m^2 \omega_{\mu}A^{\mu}$ to $2m \xi_{\mu}A^{\mu}=2mA^{\mu}E^{\alpha\beta}\omega_{\beta(\alpha\mu)}$ in such a way that the link is now invariant under the gauge transformation (\ref{gauge1}). Last, the dual map between the equations of motion which comes from the second order action (\ref{EHCS1})  and the equations of motion of the first order action (\ref{sd1}) can be obtained through the dual field $F^{\mu(\beta\gamma)}$:

\be \omega^{\mu(\beta\gamma)}\longleftrightarrow F^{\mu(\beta\gamma)}(\omega,A)=\frac{\Omega^{\mu(\beta\gamma)}(\xi)}{m}+f^{\mu(\beta\gamma)}(A).\label{dualmap1}\ee

\no where $f^{\mu(\beta\gamma)}(A)$ is defined in (\ref{fa}). With this dual map, one can reproduce the equations of motion of the first order self-dual model from the equations of motion of the second order self-dual model. Note that $F^{\mu(\beta\gamma)}$ is gauge invariant.

We point out here that the new auxiliary lagrangian (\ref{lagaux1}), has been automatically generated through the $NGE$ procedure as well as the gauge invariant source term for the auxiliary ghost-killing field $A_{\mu}$. In the next section we show that it is possible to continue with the $NGE$ procedure in order to obtain a third order self dual model.

\subsection{From $SD(2)$ to $SD(3)$}
\subparagraph{} 
The second order self-dual model obtained before is invariant under the gauge symmetry (\ref{gauge1}). As we have done for spin-2 \cite{ddelm1} we are going to impose a new gauge symmetry. One can do this by generalizing the gauge symmetry used in the spin-2 case \footnote{In that case the symmetry corresponds to an arbitrary shift in the antisymmetric part of the rank-2 tensor $\delta e_{\mu\nu}=\Lambda_{[\mu\nu]}$} let us propose the following symmetry:
\be \delta_{\Phi}\omega_{\mu(\beta\gamma)}=\epsilon_{\mu\beta}^{\s\s\rho}\Phi_{(\rho\gamma)}+\epsilon_{\mu\gamma}^{\s\s\rho}\Phi_{(\rho\beta)}.\label{gauge2}\ee

\no It is straightforward to verify that the second order term in (\ref{EHCS1}) is invariant under (\ref{gauge2}). However the first order Chern-Simons like term breaks this symmetry. Then one can use again the $NGE$ procedure. From (\ref{EHCS1}) we calculate the new Euler tensor, which is now of second order in derivatives:

\be \frac{\delta S_{SD(2)}}{\delta \omega_{\mu(\beta\gamma)}}= L^{\mu(\beta\gamma)}= E^{\mu}_{\s\lambda}\left(\Omega^{\lambda(\beta\gamma)}(\xi)+m\,\omega^{\lambda(\beta\gamma)}+f^{\lambda(\beta\gamma)}(A)-\frac{1}{m}\Omega^{\lambda(\beta\gamma)}(j)  \right)
\equiv E^{\mu}_{\s\lambda}\tilde{L}^{\lambda(\beta\gamma)}
.\label{euler2}\ee

\no Following the NGE procedure, we propose the first iteration given by:
\be S_1=S_{SD(2)}-\int d^3x\,\,b_{\mu(\beta\gamma)}L^{\mu(\beta\gamma)},\label{s22}\ee

\no were $b_{\mu(\beta\gamma)}$ is an auxiliary field with the same symmetry properties of $\omega_{\mu(\beta\gamma)}$ i.e; $\delta_{\Phi} b_{\mu(\beta\gamma)}=\delta_{\Phi}\omega_{\mu(\beta\gamma)}$. Then, with respect to the gauge transformation (\ref{gauge2}) the action (\ref{s22}) has the following gauge transformation:

\be \delta_{\Phi}S_1=-\frac{m}{2}\int d^3x\,\,\delta_{\Phi}\,(b_{\mu(\beta\gamma)}E^{\mu}_{\s\lambda}b^{\lambda(\beta\gamma)}).\ee

\no Then,by construction we have the following action:
\be S_2=S_{SD(2)}-\int d^3x\,\,\left(b_{\mu(\beta\gamma)}L^{\mu(\beta\gamma)}-\frac{m}{2}b_{\mu(\beta\gamma)}E^{\mu}_{\s\lambda}b^{\lambda(\beta\gamma)}  \right),\ee

\no which is automatically invariant under (\ref{gauge2}) and also under (\ref{gauge1}). Note that we can rewrite $S_2$  in the following way:

\be S_2= S_{SD(2)}+\int d^3x\,\,\left\lbrack\frac{m}{2}\left( b_{\mu(\beta\gamma)}-\frac{\tilde{L}_{\mu(\beta\gamma)}}{m}\right)E^{\mu}_{\s\lambda}
\left(b^{\lambda(\beta\gamma)}-\frac{\tilde{L}^{\lambda(\beta\gamma)}}{m}\right)\right.
-\left.\frac{1}{2m}\tilde{L}_{\mu(\beta\gamma)}E^{\mu}_{\s\lambda}\tilde{L}^{\lambda(\beta\gamma)} \right\rbrack,\label{desacopla}\ee

\no where $\tilde{L}^{\lambda(\beta\gamma)}$ is defined in (\ref{euler2}). Making the change of variable $b^{\lambda(\beta\gamma)}\to \tilde{b}^{\lambda(\beta\gamma)}+\tilde{L}^{\lambda(\beta\gamma)}/m$, the first term in (\ref{desacopla}) gets decoupled. The term $m \tilde{b}_{\mu(\beta\gamma)}E^{\mu}_{\s\lambda}\tilde{b}^{\lambda(\beta\gamma)}/2$ has no particle content, then we have:

\be S_2= S_{SD(2)}-\frac{1}{2m}\int d^3x\,\,\tilde{L}_{\mu(\beta\gamma)}E^{\mu}_{\s\lambda}\tilde{L}^{\lambda(\beta\gamma)}.\label{3.46}\ee

\no Note that the the equations of motion $E^{\mu}_{\s\lambda}\tilde{L}^{\lambda(\beta\gamma)}=L^{\mu(\beta\gamma)}$ of the second order self-dual model $S_{SD(2)}$, i.e. $L^{\mu(\beta\gamma)}=0$ are embedded in the equations of motion of $S_2$. Substituting $\tilde{L}_{\mu(\beta\gamma)}$ in (\ref{3.46}) we have after some manipulation the third-order self-dual model:

\bea S_{SD(3)} &=& \int d^3x\,\,\Bigg[ -\frac{1}{2}\xi_{\mu(\beta\gamma)}\Omega^{\mu(\beta\gamma)}(\xi)-\frac{1}{2m}\Omega_{\mu(\beta\gamma)}(\xi)E^{\mu}_{\s\lambda}\Omega^{\lambda(\beta\gamma)}(\xi)\nn\\
 &-&  f_{\mu(\beta\gamma)}(A)E^{\mu}_{\s\lambda}\Omega^{\lambda(\beta\gamma)}(\xi)-j_{\mu(\beta\gamma)}H^{\mu(\beta\gamma)}(\omega,A)\Bigg] + S^{''}[A],\label{s13}\eea

\no which is precisely the third order ``topologically'' massive spin-3 action proposed by Deser and Damour in \cite{deserdam}. The action is invariant under the gauge symmetries (\ref{gauge1}) and (\ref{gauge2}). The second term in (\ref{s13}) corresponds to the ``topologically'' Chern-Simons term of third order in derivatives.  As before the auxiliary lagrangian has automatically changed in order to get rid of lower spin ghosts:
\be S^{''}[A]=\int d^3x\,\,\left\lbrack-\frac{32 m}{3}\,\,\epsilon^{\mu\nu\alpha}A_{\mu}\p_{\nu}A_{\alpha}-\frac{32m^2}{3}\,\,A_{\mu}A^{\mu}
\,-12(\p_{\mu}A^{\mu})^2 \right\rbrack.\label{lagaux3}\ee 

The auxiliary lagrangian (\ref{lagaux3}), see also \cite{deserdam} now differs from the first one (\ref{lagaux1}) by two numerical factors, by this time the Chern-Simons term is also modified. One can notice that this modification corresponds to the same numerical factor which has appeared before when we found (\ref{lagaux2}). Again one can observe that the linking term (the third term in (\ref{s13})) between the spin-3 fields and the auxiliary vector field is modified, becoming invariant under the new gauge symmetry (\ref{gauge2}). Last, in this case the equivalence between the third order self-dual model $S_{SD(3)}$ and $S_{SD(2)}$ is given by the dual map given by the dual field $H^{\mu(\beta\gamma)}$ coupled to the  source term in (\ref{s13}):

\be \omega_{\mu(\beta\gamma)}\longleftrightarrow H^{\mu(\beta\gamma)}=-\frac{1}{m}\Omega^{\mu(\beta\gamma)}\left(\frac{\Omega}{m}+f\right)+f^{\mu(\beta\gamma)}(A).\label{dual3}\ee

The third order theory that we have found here is expressed in terms of the partially symmetric field $\omega_{\mu(\beta\gamma)}$ (vierbein-like formulation), however one can compare our result with the ``topologically'' massive spin-3 theory \cite{deserdam} which is given in terms of totally symmetric fields, by doing the following change of variables:

\be \omega_{\mu(\beta\gamma)}= \frac{1}{\sqrt{3}}\left[\phi_{\mu\beta\gamma}+\frac{1}{4}(\eta_{\lambda\beta}\phi_{\gamma}+\eta_{\lambda\gamma}\phi_{\beta})-\frac{1}{2}\eta_{\beta\gamma}\phi_{\lambda}\right]+ (\epsilon_{\mu\nu\beta}\chi^{\nu}_{\s\gamma}+\epsilon_{\mu\nu\gamma}\chi^{\nu}_{\s\beta}),\label{deco}\ee

\no where  $\chi_{\mu\nu}(x)=\chi_{\nu\mu}(x)$ and $\eta^{\mu\nu}\chi_{\mu\nu}=\chi=0$. Even if we had $\chi\neq 0$, (\ref{deco}) would be invariant under a Weyl transformation, in other words $\delta_{\varphi} \chi_{\mu\nu}=\eta_{\mu\nu}\varphi$. The numerical factors in (\ref{deco}) are obtained in such a way that our results fit the results of  \cite{deserdam}. In $D=2+1$, the spin-3 basic field $\omega_{\mu(\beta\gamma)}$ has $15$  independent components. This can be verified by noticing that $\omega_{\mu(\beta\gamma)}$ has by definition the number of independent components of a vector, times the number of independent components of a symmetric traceless rank two tensor. On the right hand side of (\ref{deco}), we have the number of independent components of a rank three symmetric tensor $\phi_{\mu\nu\lambda}$, which is $D(D+1)(D+2)/6=10$ in $D=2+1$ plus the number of independent components of a traceless rank two symmetric tensor $\chi_{\mu\nu}$ which is $D(D+1)/2-1=5$ in $D=2+1$. Then in three dimensions we have $15$ independent components on both sides of (\ref{deco}).

Rewriting (\ref{s13}) in terms of totally symmetric fields we obtain:

\bea S_{SD(3)}[\phi,A]&=&\int d^3x\,\,\left\lbrack -\frac{1}{2}\phi_{\mu\beta\gamma} G^{\mu\beta\gamma}(\phi)-\frac{1}{2m} C_{\mu\beta\gamma}(\phi)G^{\mu\beta\gamma}(\phi)-\frac{4}{3\sqrt{3}}\tilde{A}_{\mu\beta\gamma}G^{\mu\beta\gamma}(\phi)\right.\nn\\
&-&\left.\frac{1}{m^2}C_{\mu\beta\gamma}(\tilde{j})G^{\mu\beta\gamma}(\phi)\right\rbrack+ S''[A] -\frac{\sqrt{3}}{m}\int \,d^3x\, \tilde{j}_{\mu}(E^{\mu\nu}A_{\nu}+mA^{\mu}).\label{s3final}\eea

\no We have used the spin-3 ``Einstein tensor'' $G_{\mu\beta\gamma}(\phi)$ and the symmetrized curl  $C_{\mu\nu\lambda}(\phi)$ defined in \cite{deserdam}, given by:

\be G^{\mu\beta\gamma}(\phi)\equiv R^{\mu\nu\lambda}-\frac{1}{2}\eta^{(\mu\nu}R^{\lambda)}\ee

\no where we have the ``Ricci'' tensor given by  $R^{\mu\nu\lambda}=\Box \phi^{\mu\nu\lambda}-\p_{\alpha}\p^{(\mu}\phi^{\alpha\nu\lambda)}+\p^{(\mu}\p^{\nu}\phi^{\lambda)}$ and its trace $R^{\lambda}=\eta_{\mu\nu}R^{\mu\nu\lambda}$. The symmetrized curl is defined by:
\be C_{\mu\beta\gamma}(\phi)\equiv E_{(\mu}^{\s\s\nu}\phi_{\nu\beta\gamma)}.\ee

\no Besides the above definitions we have used the symmetric combinations for the spin-1 field $A_{\mu}$: 

\be \tilde{A}_{\mu\nu\lambda}=A_{(\mu}\eta_{\nu\lambda)}\ee  

\no and  for the source term of (\ref{s13}):

\be j_{\mu(\beta\gamma)}=\frac{1}{\sqrt{3}}\left\lbrack \tilde{j}_{\mu\beta\gamma}+\frac{1}{4}(\eta_{\mu\beta}\tilde{j}_\gamma+\eta_{\mu\gamma}\tilde{j}_{\beta})-\frac{1}{2}\eta_{\beta\gamma}\tilde{j}_{\mu}\right\rbrack.\ee

\no It is useful for the next step to notice that the first two terms in (\ref{s3final}) are self-adjoint, i.e; $\phi_{\mu\nu\lambda}G^{\mu\nu\lambda}(\psi)=\psi_{\mu\nu\lambda}G^{\mu\nu\lambda}(\phi)$ and $\phi_{\mu\nu\lambda}C^{\mu\nu\lambda}(\psi)=\psi_{\mu\nu\lambda}C^{\mu\nu\lambda}(\phi)$ inside spacetime integrals.  From this totally symmetric version of the third order self-dual model, in the next section we are going to make the last step with the $NGE$ procedure by imposing a new gauge symmetry on (\ref{s3final}). Notice that $\chi^{\mu}_{\s\nu}$ introduced in (\ref{deco}) is absence in (\ref{s3final}) due to the symmetry (\ref{gauge2}).

\subsection{A complete fourth order self-dual action for spin-3}
\subparagraph{}
In the first section we have used the gauge symmetry (\ref{gauge1}) where the symmetric rank-2 parameter $\tilde{\Lambda}_{\mu\nu}$ is traceless. However one can verify that the spin-3 topological Chern-Simons term of third order, second term in (\ref{s3final}), is invariant with respect to a generalization of this symmetry with an arbitrary (traceful) symmetric parameter:

\be \delta_{\Lambda}\phi_{\mu\beta\gamma}=\p_{(\mu}\Lambda_{\beta\gamma)},\label{gauge4}\ee

\no with $\Lambda_{\beta\gamma}=\Lambda_{\gamma\beta}$. On the other hand the second order term, first term in (\ref{s3final}) as well as the interaction term between the vector field $A_{\mu}$ and the symmetric spin-3 field $\phi_{\mu\nu\lambda}$ are non invariant under the generalization (\ref{gauge4}). So one can now impose this symmetry in one more round of the $NGE$ procedure.  By noticing that the Einstein operator $G_{\mu\beta\gamma}(\phi)$ is self-adjoint, the $\phi_{\mu\beta\gamma}$ variation of the action (\ref{s3final}) gives us  the following Euler tensor:

\be \frac{\delta S_{SD(3)}}{\delta \phi_{\mu\nu\lambda}}\equiv N^{\mu\nu\lambda}=-G^{\mu\nu\lambda}\left\lbrack \phi+\frac{C(\phi)}{m}+\frac{4\tilde{A}}{3\sqrt{3}}+\frac{C(\tilde{j})}{m^2}\right\rbrack\equiv -G^{\mu\nu\lambda}(b),\label{Ks}\ee

\no where we have automatically defined $b$. As usual we start with a first iteration of the form:

\be S_1=S_{SD(3)}-\int d^3x\s a_{\mu\nu\lambda}N^{\mu\nu\lambda}\label{s14}\ee

\no where we have added a totally symmetric field $a_{\mu\beta\gamma}$ such that its gauge transformation is given by $\delta_{\Lambda} a_{\mu\nu\lambda}=\p_{(\mu}\Lambda_{\nu\lambda)}=\delta_{\Lambda}\phi_{\mu\nu\lambda}$. Then, from the gauge transformation of (\ref{s14}) we have:

\be \delta_{\Lambda} S_1= \int d^3x\s a_{\mu\nu\lambda}\delta_{\Lambda} G^{\mu\nu\lambda}(a),\ee

\no which then gives us:

\be S_2= S_{SD(3)}-\int d^3x\s \left\lbrack- a_{\mu\nu\lambda}G^{\mu\nu\lambda}(b)+\frac{1}{2}a_{\mu\nu\lambda}G^{\mu\nu\lambda}(a)\right\rbrack.\label{s2}\ee

\no where $b$ is defined in (\ref{Ks}). This allows us to rewrite (\ref{s2}) as:

\be S_2= S_{SD(3)}- \int d^3x \left\lbrack \frac{1}{2}(a_{\mu\nu\lambda}-b_{\mu\nu\lambda})G^{\mu\nu\lambda}(a-b)-\frac{1}{2}b_{\mu\nu\lambda}G^{\mu\nu\lambda}(b)\right\rbrack.\ee 

\no Finally, shifting $a_{\mu\nu\lambda}\to \tilde{a}_{\mu\nu\lambda}+b_{\mu\nu\lambda}$ we can decouple $\tilde{a}_{\mu\nu\lambda}$ and $b_{\mu\nu\lambda}$ and since the second order term $\tilde{a}_{\mu\nu\lambda}G^{\mu\nu\lambda}(\tilde{a})$ is completely decoupled and has no particle content we end up with the equivalent invariant action:

\be S_2= S_{SD(3)}+\frac{1}{2}\int \,d^3x\,\,b_{\mu\nu\lambda}G^{\mu\nu\lambda}(b).\ee

\no Substituting back $b_{\mu\nu\lambda}$ given in (\ref{Ks}) we have a complete fourth-order action given by:

\bea S_{SD(4)}&=&\int d^3x\s \left\lbrace \frac{1}{2m}\phi_{\mu\nu\lambda}G^{\mu\nu\lambda}\left\lbrack C(\phi)\right\rbrack +\frac{1}{2m^2}C_{\mu\nu\lambda}(\phi)G^{\mu\nu\lambda}\left\lbrack C(\phi)\right\rbrack +\frac{4}{3\sqrt{3}m}\tilde{A}_{\mu\nu\lambda}G^{\mu\nu\lambda}\left\lbrack C(\phi)\right\rbrack\right.\nn\\
&+&\left. \frac{1}{m^2}C_{\mu\nu\lambda}(\tilde{j})G^{\mu\nu\lambda}\left\lbrack C(\phi)+\frac{4}{3\sqrt{3}m}\tilde{A}\right\rbrack \right\rbrace+ S'''[A]-\frac{\sqrt{3}}{m}\, \int d^3x\,\,\tilde{j}_{\mu}\,\,(E^{\mu\rho} A_{\rho}+mA^{\mu})\nn\\ \label{sd4}\eea

Now the auxiliary action $S^{'''}[A]$ has gained a new term of second order in derivatives, which combined with $(\p_{\mu}A^{\mu})^2$,  is precisely the Maxwell term:

\be S'''[A]= -\frac{32}{3}\int d^3x\,\,\left\lbrack -\frac{1}{2}F_{\mu\nu}F^{\mu\nu}+m\epsilon^{\mu\nu\alpha}A_{\mu}\p_{\nu}A_{\alpha}+m^2 A_{\mu}A^{\mu}\right\rbrack \ee

\no where $F_{\mu\nu}=\p_{\mu}A_{\nu}-\p_{\nu}A_{\mu}$. In the literature there is no fourth order action describing a massive spin-3 singlet in $D=2+1$. In \cite{tow3} equations of motion of fourth order in derivatives are introduced but there is no an action from where one can derive it. Besides, the auxiliary fields needed for a complete description of the spin three parity singlet without ghosts are not considered. Here we are introducing a complete model with an action for the auxiliary fields. And the equivalence, is guaranteed through the dual maps. We can check that the spin-3 sector of (\ref{sd4}) is indeed the action that gives us the fourth order equations of motion of \cite{tow3} by rewriting the first two terms in the action as follow:

\be \int d^3x\s  \frac{1}{2m}\phi_{\mu\nu\lambda}G^{\mu\nu\lambda}\left\lbrack C(\phi)\right\rbrack= -\frac{3}{2m} \int d^3x\s \phi^{\mu\nu\lambda}E_{\mu}^{\s \alpha}E_{\nu}^{\s \beta}E_{\lambda}^{\s\gamma}\phi_{\alpha\beta\gamma} \ee 

\be \int d^3x\s  \frac{1}{2m^2}C_{\mu\nu\lambda}(\phi)G^{\mu\nu\lambda}\left\lbrack C(\phi)\right\rbrack= \frac{9}{2m^2} \int d^3x \s \phi^{\mu\nu\lambda}\Box \theta_{\mu}^{ \alpha}\,\,E_{\nu}^{\s \beta}\,\,E_{\lambda}^{\s\gamma}\phi_{\alpha\beta\gamma} \ee

\no The combination of these two terms according to (\ref{sd4}) allows us to derive  derive the equations of motion suggested in \cite{tow3}. In that paper the authors have defined a 3rd rank (and 3rd order) symmetric tensor potential which is basically a symmetric combination of the operators $E_{\mu\nu}$ that we have used along this paper. We must say that they have considered as the Einstein tensor this 3rd rank tensor instead of the definition used in \cite{deserdam} which is of second order in derivatives and this is not clear at nonlinear level . Throughout  this work we have preferred to keep the second order definition of the spin-3 analogue of the Einstein tensor.
\section{Conclusion}

We have verified that there is an equivalence between four self dual models for massive spin-3 particles in $D=2+1$. We have done this through the $NGE$ procedure. The same procedure was used before in the context of massive spin-2 and spin-1 self-dual models. The challenges here were the presence of auxiliary Lagrangians which have to be considered in order to preserve only spin-3 propagations without ghosts and the identification of the correct symmetry to be embedded. We have observed that the auxiliary Lagrangians have been automatically generated by the $NGE$ procedure . The changes  guarantees the absence of lower spin ghosts.

Although we have started with the first order self-dual model \cite{AragoneS31} passing through the second order self-dual model \cite{AragoneS32} in the vierbein formulation, one could verify that after arriving in the ``topologically'' massive spin-3 model it is possible to make a change of variables which relates the partially symmetric formulation $\omega_{\mu(\beta\gamma)}$ and the totally symmetric formulation $\phi_{\mu\beta\gamma}$. Such change of variables see (\ref{deco}) preserves, and respect the number of independent degrees of freedom.

A complete fourth order action was achieved in (\ref{sd4}). The core of this action reproduces the fourth order spin-3 equations of motion proposed by \cite{tow3}. However in \cite{tow3}  the auxiliary ghost-killing fields needed in order to maintain the correct spin-3 propagation have not been considered. Here the auxiliary action is obtained systematically from previous massive spin-3 self-dual models once we know the first order model (\ref{sd1}).

We have used a generalization of the symmetries used in the spin-2 context up to the third order model. However to obtain the fourth order self-dual model a generalization of the Weyl transformation for spin-3, which in our point of view would be $\delta \phi_{\mu\beta\gamma}=\eta_{(\mu\beta}\xi_{\gamma)}$ do not correspond to the necessary symmetry to make the last step. Instead of this we have generalized the symmetry used between the first and second order self-dual model, taking advantage that the ``topological'' Chern-Simons term is invariant under $\delta \phi_{\mu\beta\gamma}=\p_{(\mu}\Lambda_{\beta\gamma)}$ with $\Lambda_{\mu\nu}$ arbitrary symmetric tensor. In the table bellow the reader can find a summary of the spins, $1$, $2$ and $3$ chains of embeddings and the corresponding symmetry and constraint on the symmetry parameters.

The next challenge is to go beyond the 4th order self-dual model $S_{SD(4)}^{(3)}$ and arrive at the 6th order self-dual model suggested in \cite{tow3} eventually. We have not been able to find any new local symmetry of the 4th order term in (\ref{s3final}) to be embedded. This is under investigation.

{\centering{\begin{table}[h!]

\centering{
\begin{tabular}{|c|c|c|c|}
\hline \rule[-2ex]{0pt}{5.5ex} {$\bf{s}$} & \bf{Embedding} & \bf{Symmetry} &  \bf{Constraint}\\ 
\hline \rule[-2ex]{0pt}{6.5ex} $1$ & $SD_{(1)}^{(1)}\rightarrow SD_{(2)}^{(1)}$ & $\delta_{\Lambda}f_{\mu}=\p_{\mu}\Lambda$ & ||\\ 
\hline \rule[-2ex]{0pt}{5.5ex} $2$ & $SD_{(1)}^{(2)}\rightarrow SD_{(2)}^{(2)}$ & $\delta_{\xi}e_{\mu\nu}=\p_{\mu}{\xi}_{\nu}$  & || \\ 
\hline \rule[-2ex]{0pt}{5.5ex} $2$ & $SD_{(2)}^{(2)}\rightarrow SD_{(3)}^{(2)}$ & $\delta_{\Lambda}e_{\mu\nu}=\Lambda_{[\mu\nu]}=\epsilon_{\mu\nu\alpha}\xi^{\alpha}$  & $\Lambda_{(\mu\nu)}=0$ \\ 
\hline \rule[-2ex]{0pt}{5.5ex} $2$ & $SD_{(3)}^{(2)}\rightarrow SD_{(4)}^{(2)}$ & $\delta_{\phi}e_{\mu\nu}=\phi\eta_{\mu\nu}$  & || \\ 
\hline \rule[-2ex]{0pt}{5.5ex} $3$ & $SD_{(1)}^{(3)}\rightarrow SD_{(2)}^{(3)}$ & $\delta_{\tilde{\Lambda}}\omega_{\mu(\beta\gamma)}=\p_{\mu}\tilde{\Lambda}_{(\beta\gamma)}$ & $\eta^{\beta\gamma}\tilde{\Lambda}_{(\beta\gamma)}=0$ \\ 
\hline \rule[-2ex]{0pt}{5.5ex} $3$ & $SD_{(2)}^{(3)}\rightarrow SD_{(3)}^{(3)}$ & $\delta_{\Phi}\omega_{\mu(\beta\gamma)}=\epsilon_{\mu\beta}^{\s\s\rho}\phi_{(\rho\gamma)}+\epsilon_{\mu\gamma}^{\s\s\rho}\Phi_{(\rho\beta)}$ & ||  \\ 
\hline \rule[-2ex]{0pt}{5.5ex} $3$ & $SD_{(3)}^{(3)}\rightarrow SD_{(4)}^{(3)}$ & $\delta_{\Lambda}\phi_{\mu\beta\gamma}=\p_{(\mu}\Lambda_{\beta\gamma)}$ & $\Lambda_{\beta\gamma}=\Lambda_{\gamma\beta}$ \\ 
\hline 
\end{tabular} }\\
\bigskip
\centering{\caption{Here we have used $SD_{n}^{s}$ where $n$ is the order of the self-dual model and $s$ is the spin.}}
\end{table}}}

\section{Appendix-A: Equations of motion and dual map}

\subsection{Equations of motion}
We start by deriving from (\ref{sd1}) with $j_{\mu(\beta\gamma)}=0$, the equations of motion with respect to the field $\omega_{\mu(\beta\gamma)}$, which gives us:
     
     \bea K^{\mu(\beta\gamma)}&=&-mE^{\mu\alpha}\omega_{\alpha}^{\s(\beta\gamma)}+\frac{m^2}{6}(\eta^{\mu\beta}\omega^{\gamma}+\eta^{\mu\gamma}\omega^{\beta}-\omega^{\beta(\mu\gamma)}-\omega^{\gamma(\mu\beta)})\nn\\
     &+&\frac{m^2}{2}\left(\eta^{\mu\beta}A^{\gamma}+\eta^{\mu\gamma}A^{\beta}-\frac{2}{3}\eta^{\beta\gamma}A^{\mu}\right)=0\label{tensor}\eea 
     
 \no and with rescpect to the vector field $A^{\mu}$:
     \be F^{\mu}= 18mE^{\mu\alpha}A_{\alpha}-18m^2A^{\mu}+m^2\omega^{\mu}+24\p^{\mu}\p_{\alpha}A^{\alpha}=0.\label{vetor}\ee

\no Our goal is then to demonstrate that from the equations of motion (\ref{tensor}) and (\ref{vetor}) one can obtain the Fierz-Pauli conditions for spin-3. In order to demonstrate it we first make some manipulations with $K_{\mu(\beta\gamma)}$, for example, applying $\p_{\mu}$ in $K^{\mu(\beta\gamma)}$ we have:
     \be \p^{\beta}\omega^{\gamma}+\p^{\gamma}\omega^{\beta}-\p_{\mu}\left(\omega^{\beta(\gamma\mu)}+\omega^{\gamma(\beta\mu)}\right)=2\eta^{\beta\gamma}\p_{\mu}A^{\mu}-3(\p^{\beta}A^{\gamma}+\p^{\gamma}A^{\beta}).\label{2}\ee
\no Now, let us take the following combination: $\epsilon_{\lambda\mu\beta}K^{\mu(\beta}_{\s\s\s\gamma)}+\epsilon_{\gamma\mu\beta}K^{\mu(\beta}_{\s\s\s\lambda)}=0$. This leave us with the equation:
     \be \p^{\beta}\omega^{\gamma}+\p^{\gamma}\omega^{\beta}-\p_{\mu}(\omega^{\beta(\mu\gamma)}+\omega^{\gamma(\mu\beta)})=\frac{m}{6}(\epsilon^{\beta}_{\s\mu\alpha}\omega^{\alpha(\mu\gamma)}+\epsilon^{\gamma}_{\s\mu\alpha}\omega^{\alpha(\mu\beta)}).\label{3}\ee
     
\no Taking the trace $\eta_{\mu\beta}K^{\mu(\beta\gamma)}$, we have:
     \be E_{\alpha\mu}\omega^{\alpha(\mu\gamma)}+\frac{m}{2}\omega^{\gamma}+\frac{5m}{3}A^{\gamma}=0.\label{4}\ee

\no Applying $\p_{\beta}\p_{\gamma}$ in (\ref{2}) and (\ref{3}) and then taking the difference between the resulting equations we have:
    \be -4\Box\p_{\mu}A^{\mu}=-\frac{m}{3}E_{\alpha\mu}\p_{\lambda}\omega^{\alpha(\mu\lambda)}.\label{boxdiv}\ee
     
\no Applying  $\p_{\gamma}$ in (\ref{4}) and using the result (\ref{boxdiv}) we find that:
     \be \Box\p_{\mu}A^{\mu}=-\frac{m^2}{24}\p_{\mu}\omega^{\mu}-\frac{5m^2}{36}\p_{\mu}A^{\mu},\label{da}\ee
    
\no Now if we apply $\p_{\mu}$ in (\ref{vetor}) we have:
    \be \left(\Box-\frac{3m^2}{4}\right)\p_{\mu}A^{\mu}=-\frac{m^2}{24}\p_{\mu}\omega^{\mu}.\label{dome}\ee
   
Using (\ref{dome}) in (\ref{da}) we conclude that $\p_{\mu}A^{\mu}=0$ and consequently $\p_{\mu}\omega^{\mu}=0$. Now let us define the following vectors $S_{\alpha}=\p_{\beta}\p_{\gamma}\omega_{\alpha}^{\s(\beta\gamma)}$ and $T_{\alpha}=\p^{\mu}\p^{\lambda}\omega_{\mu(\lambda\alpha)}$. If we make $\p_{\beta}\p_{\gamma}K^{\mu(\beta\gamma)}$ and $\p_{\mu}\p_{\beta}K^{\mu(\beta\gamma)}$ we have respectively:
    
    \be E^{\mu\alpha}S_{\alpha}+\frac{m}{3}(T^{\mu}+\Box A^{\mu})=0,\label{V1}\ee
    
    \be \Box \omega^{\gamma}+3\Box A^{\gamma}=T^{\gamma}+S^{\gamma}.\label{V2}\ee
    
 Then taking $E_{\mu\beta}K^{\mu(\beta\gamma)}$, after some algebra it is possible to show that:
    \be T^{\gamma}=\Box\omega^{\gamma}-3\,\,\Box A^{\gamma}+\frac{16m}{3}E^{\gamma\beta}A_{\beta}-\frac{16m^2}{9}A^{\gamma},\label{V3}\ee
    
\no so, from (\ref{V3}) and (\ref{V2}) we have $S^{\gamma}$ written as: 
    \be S^{\gamma}=6\,\,\Box A^{\gamma}-\frac{16m}{3}E^{\gamma\beta}A_{\beta}+\frac{16m^2}{9}A^{\gamma}.\label{V3b}\ee
    
\no Applying $E^{\beta\gamma}$ in (\ref{4}) we have:
    \be S^{\beta}-T^{\beta}+\Box\omega^{\beta}+\frac{32m}{3}E^{\beta\gamma}A_{\gamma}-9\Box A^{\beta}=0.\label{V4}\ee
    
Finally substituting back (\ref{V3}) and (\ref{V3b}) in (\ref{V4}), after some manipulation we can prove that   
     $A_{\gamma}=0$, and consequently from (\ref{vetor}) we can demonstrate that $\omega_{\gamma}=0$. Back with this results in (\ref{V2}) and (\ref{V4}) we show that $T^{\mu}=0=S^{\mu}$. Besides, making $A_{\gamma}=0=\omega_{\gamma}$ in (\ref{2}) we can verify that $\omega_{\mu(\beta\gamma)}$ obey:
    \be \p_{\mu}(\omega^{\beta(\gamma\mu)}+\omega^{\gamma(\beta\mu)})=0.\ee
     
     \no So, from (\ref{3}) the antissimetric part of $\omega_{\mu(\beta\gamma)}$ is null, i.e.:
        
        \be \omega^{\mu(\beta\gamma)}-\omega^{\beta(\mu\gamma)}=0.\ee
       
    Using this new informations, after applying $E_{\lambda\mu}$ in (\ref{tensor}) we can check that:
    \be \left(\Box-\frac{m^2}{9}  \right) \omega_{\lambda}^{\s(\beta\gamma)}=0.\ee

    \no Here, one could then rescale the mass by a factor $3$, however we prefer to keep the same notation of \cite{AragoneS31}. Back with these results in (\ref{4}), and applying $\p_{\gamma}$ in  (\ref{2}) we can demonstrate that $\omega_{\mu(\beta\gamma)}$ is transverse $\p_{\gamma}\omega^{\mu(\beta\gamma)}=0$.
   \no Summarizing we have demonstrated that all the Fierz-Pauli conditions are satisfied from the equations of motion, i.e., the spin-3 field is traceless $\omega_{\gamma}=0$, totally symmetric, transverse and satisfy a Klein-Gordon equation. It is a good moment to also verify that from the equations of motion one can obtain the Pauli-Lubanski equation, which makes clear that the spin described is in fact $3$. In order to make this we need to define the generators of translation and rotation for spin-3 states. The general expression for the generator of rotations can be obtained, for example, from \cite{arias}. For the reader convenience we present here the explicit form of the $s=3$ case:
    \bea (J_3^{\alpha})^{\beta\gamma\lambda}_{\mu\nu\rho}&=&\frac{i}{12}\epsilon_{(\mu}^{\s\alpha(\beta}\delta^{\gamma}_{\nu}\delta^{\lambda)}_{\rho)}\nn\\
    &=&\frac{i}{3}\left(\epsilon_{\mu}^{\s\alpha\beta}{\cal{I}_{S}}^{\gamma\lambda}_{\nu\rho}+\epsilon_{\nu}^{\s\alpha\beta}{\cal{I}_{S}}^{\gamma\lambda}_{\mu\rho}+\epsilon_{\rho}^{\s\alpha\beta}{\cal{I}_{S}}^{\gamma\lambda}_{\nu\mu}\right.\nn\\
    &+& \left.\epsilon_{\mu}^{\s\alpha\gamma}{\cal{I}_{S}}^{\beta\lambda}_{\nu\rho}+\epsilon_{\nu}^{\s\alpha\gamma}{\cal{I}_{S}}^{\beta\lambda}_{\mu\rho}+\epsilon_{\rho}^{\s\alpha\gamma}{\cal{I}_{S}}^{\beta\lambda}_{\mu\nu} \right.\nn\\
    &+& \left.\epsilon_{\mu}^{\s\alpha\lambda}{\cal{I}_{S}}^{\beta\gamma}_{\nu\rho}+\epsilon_{\nu}^{\s\alpha\lambda}{\cal{I}_{S}}^{\beta\gamma}_{\mu\rho}+\epsilon_{\rho}^{\s\alpha\lambda}{\cal{I}_{S}}^{\beta\gamma}_{\mu\nu}\right),\label{J3}\eea

    \no where:
    \be {\cal{I}_{S}}_{\mu\nu}^{\beta\gamma}=\frac{(\delta_{\mu}^{\beta}\delta_{\nu}^{\gamma}+\delta_{\nu}^{\beta}\delta_{\mu}^{\gamma})}{2},\ee
    
    \no is the rank-2 symmetric identity. The generator of rotations $J_3$ given by (\ref{J3}) obey the following relation:
    \be (J_3^{\alpha})^{\beta\gamma\lambda}_{\mu\nu\rho}(J_3^{\alpha})^{\sigma\phi\omega}_{\beta\gamma\lambda}=s(s+1){\cal{I}}^{\sigma\phi\omega}_{\mu\nu\rho},\label{JJI}\ee
    
    \no with $s=3$ in this case. In (\ref{JJI}) we have:
    \be {\cal{I}}^{\sigma\phi\omega}_{\mu\nu\rho}=-{\cal{I}_{S}}^{\sigma\phi\omega}_{\mu\nu\rho}+\frac{1}{18}\left[\delta_{\mu\nu}(\delta_{\rho}^{\phi}\delta^{\omega\sigma}+\delta^{\sigma\phi}\delta_{\rho}^{\omega}+\delta_{\rho}^{\sigma}\delta^{\omega\phi})
    +(\rho \leftrightarrow \nu)
    +(\mu \leftrightarrow \rho)\right].\ee

    \no where:
    \be {\cal{I}_{S}}_{\mu\nu\rho}^{\sigma\phi\omega}=\frac{1}{3}(\delta_{\mu}^{\sigma}{\cal{I}_{S}}_{\nu\rho}^{\phi\omega}+\delta_{\mu}^{\phi}{\cal{I}_{S}}_{\nu\rho}^{\sigma\omega}+\delta_{\mu}^{\omega}{\cal{I}_{S}}_{\nu\rho}^{\sigma\phi}),\ee

    \no is the rank-3 symmetric identity. Besides the relation (\ref{JJI}) we can also verify the following commutation relation :
    \be \left[(J_3^{\alpha})^{\beta\gamma\lambda}_{\mu\nu\rho},(J_3^{\xi})_{\beta\gamma\lambda}^{\sigma\phi\omega} \right]=i\epsilon_{\delta}^{\s\xi\alpha}(J_3^{\delta})^{\sigma\phi\omega}_{\mu\nu\rho}.\label{JJJ}\ee

    Finally, from (\ref{tensor}) using all Fiez-Pauli conditions we have the Pauli-Lubanski equation:
    \be \left(P\cdot J_3-sm\right)_{\mu\nu\rho}^{\sigma\phi\omega}{\omega}_{\sigma\phi\omega}=0.\ee
    
    \no Where $s=3$ and ${\omega}_{\sigma\phi\omega}$ is totally symmetric and traceless.
    
    \subsection{Dual map}
    From (\ref{EHCS1}) with ($j_{\mu(\beta\gamma)}=0$) we take the equations of motion for $\omega^{\mu(\beta\gamma)}$, which can be written as:
    
    \be -mE^{\mu}_{\s\alpha}F^{\alpha(\gamma\beta)}+m\xi^{\mu(\beta\gamma)}=0.\label{eq2}\ee
    
    \no where we have used the definition of the dual field $F^{\mu(\beta\gamma)}$ given by (\ref{dualmap1}). But with the same definition of $F^{\mu(\beta\gamma)}$ one can write:
    
    \bea m\xi^{\beta(\mu\gamma)}&=&-\frac{m^2}{6}(F^{\gamma(\mu\beta)}+F^{\mu(\gamma\beta)})+\frac{m^2}{6}(\eta^{\mu\beta}F^{\gamma}+\eta^{\gamma\beta}F^{\mu})\nn\\
    &+&\frac{m^2}{2}(\eta^{\mu\beta}A^{\gamma}+\eta^{\gamma\beta})-\frac{m^2}{3}\eta^{\mu\gamma}A^{\beta}.\eea
    
    \no Substituting back this result in (\ref{eq2}) we have exactly the same equations of motion derived from the first order self-dual model given by (\ref{tensor}) with the change $\omega_{\mu(\beta\gamma)} \to F_{\mu(\beta\gamma)}$. Then the equivalence between the first order self-dual model and the second order self-dual model can be demonstrated at least at the level of the equations of motion. The same procedure can be applied to demonstrate the equivalence between the other self-dual models.

\section{Acknowledgments} E.L.M and D.D thanks to Marc Henneaux for the reception of E.L.M at ULB, Brussels, Belgium where part of this work was developed. The authors thanks to  A. Khoudeir for suggestions.  D.D thanks to CNPq and FAPESP for financial support. E.L.M thanks to CNPq and CAPES for financial support.

 \end{document}